\documentstyle[amsfonts,aps,preprint]{revtex}

\newcommand{\p}{\partial}

\def\[{\left\lbrack}
\def\]{\right\rbrack}

\def\({\left(}
\def\){\right)}

\begin{document}

\title{Non-Abelian BFFT embedding, Schr\"odinger quantization and the anomaly of the $O(N)$ nonlinear sigma model}
\author{E. M. C. Abreu $^a$\thanks{Financially supported by Funda\c{c}\~ao de Amparo \`a 
Pesquisa do EstEado de S\~ao Paulo (FAPESP).}, 
J. Ananias Neto${}^{b}$, W. Oliveira${}^{b}$ and G. Oliveira-Neto${}^{b}$\thanks{Financially supported by 
Funda\c{c}\~ao de Amparo \`a Pesquisa do Estado de Minas Gerais (FAPEMIG), 
grant number 00037/99.}}
\address{${}^{a}$Departamento de F\'\i sica e Qu\'\i mica, Universidade Estadual Paulista, \\
Av. Ariberto Pereira da Cunha 333, Guaratinguet\'a, 12516-410,\\ S\~ao Paulo,
SP, Brazil, \\
E-mail: everton@feg.unesp.br\\
${}^{b}$Departamento de F\'{\i}sica, ICE,  Universidade Federal 
de Juiz de Fora, \\ Juiz de Fora, $36036-330$, Minas Gerais, MG, Brazil, \\
E-mail: jorge@fisica.ufjf.br, gilneto@fisica.ufjf.br and wilson@fisica.ufjf.br}

\date{\today}

\draft

\maketitle

\begin{abstract}
We embed the $O(N)$ nonlinear sigma model in a non-Abelian gauge theory. As a first class system, it is quantized using two different approaches: the functional Schr\"odinger method and the non-local 
field-antifield procedure. Firstly, the quantization is performed with the functional Schr\"{o}dinger method, for $N=2$, obtaining the wave functionals for the ground and excited states. In the second place, using the BV formalism we compute the one-loop anomaly. 
This important result shows that the classical gauge symmetries, appearing due to the conversion via BFFT method, are broken at the quantum level.
\end{abstract}
\vskip .5 cm
PACS number: 11.10.-z, 11.10.Ef, 11.15.-q, 11.15.Tk

\newpage

\section{Introduction }

Batalin, Fradkin, Fradkina, and Tyutin (BFFT) \cite{BFF,BT} developed
an elegant formalism for embedding second class systems in first class ones \cite{Dirac}. This is
achieved with the aid of auxiliary fields that extend the phase space in a convenient way to transform the second-class into first class constraints. 

Originally the BFFT method was formulated in a way that the first class constraints satisfy an Abelian algebra.  Banerjee {\it et al}~\cite{Banerjee1}, studying the non-Abelian Proca model, have
adapted the BFFT method in order that the first class constraints obey a non-Abelian algebra. This possibility pointed out by Banerjee {\it et al.} leads to a richer structure compared with the usual BFFT case. 
Recently, the Abelian and non-Abelian BFFT extension \cite{Banerjee1} were used to transform the $SU(2)$ Skyrme model in an Abelian and non-Abelian gauge 
theories \cite{ON}, respectively. 

In this work we use this non-Abelian extension of BFFT formalism to convert the second class constraints of the $O(N)$ nonlinear sigma model into first class ones. The corresponding Hamiltonian is derived solving a 
differential equation in an unknown function of the auxiliary fields. The Lagrangian that leads to this new theory is also derived.

The functional Schr\"{o}dinger representation
has recently been systematically used in order
to quantize different field theories, including
gravity \cite{jac,hatfield,jackiw1,gil}. Many theoretical as
well as some physical predictions have been
derived, for different theories, from the
wave-functionals obtained so far. One example of
an important theoretical feature of gauge 
theories established in the context of the
functional Schr\"{o}dinger representation,
without any `instanton' approximation, is the
so-called vacuum angle \cite{jac}. On the other
hand, from the wave-functional of the quantum
Schwarzschild-de Sitter black hole one is able
to predict how it depends on the mass and
cosmological constants \cite{gil}.

Here, we quantize the first class O(2) nonlinear sigma model using the functional Schr\"{o}dinger representation. Since this theory is constrained we apply the so-called ``reduced phase-space" quantization procedure\cite{ht}. The crucial step, in this section, is the polar transformation
from the original fields $(\phi_1,\phi_2)$ to new
fields $(R,\Theta)$. This transformation is naturally
suggested by the O(2) symmetry of the theory. In terms
of $(R,\Theta)$ the functional Schr\"{o}dinger equation is greatly simplified. From this equation it is clear
that the energy of the theory is divided in two parts: a radial one (depends only on R) and an angular one (depends only on $\Theta$). With an appropriated
suggestion for the ground state energy, we explicitly compute the ground state wave-functional
and indicate how to calculate the excited
states wave-functionals.

The field-antifield formalism, created by I. Batalin and G. Vilkovisky (BV method) \cite{BV}, has 
been used successfully to quantize the most difficult gauge theories such as supergravity theories and 
topological field theories in the Lagrangian framework \cite{GPS,Jon,Hen}. The BV method comprises the 
Faddeev-Popov quantization \cite{FP} and has the BRST symmetry as its fundamental principle 
\cite{BRST}.  The method has 
introduced the definition of the antifields which are the sources of the BRST transformation, 
i.e., for each field we have an antifield canonically conjugated in terms of the antibracket 
operation.  With the fields, the antifields and the BRST transformation we can construct the 
classical BV action.  A mathematical ingredient, called the antibracket, helps 
us to construct the fundamental equation of the formalism at the classical level, the so-called 
master equation. We may mention an extension of the BV formalism where one works in the BRST superspace.
There, the main ingredient is the definition of the superfields. The details can be found in \cite{ABD}.

At the quantum level, we can define another mathematical operator, the $\Delta$ operator, which 
is a second order differential operator. From the classical BV action and its local 
counter terms, we can construct the quantum BV action and analogously to the classical case, 
the quantum master equation.

The quantization is performed via the usual functional integration through the definition of 
the generating functional and with the help of the well known Legendre transformation with 
respect to the sources $J A$. 
When it is not possible to find a solution to the quantum master equation we can say that the 
theory has an anomaly.  The presence of a $\delta(0)$ term in the $\Delta$ operation demand a 
method to treat this divergence conveniently.  
There are various methods to regularize the theory such as Pauli-Villars\cite{Pauli}, BPHZ 
\cite{BPHZ,Jonghe} and dimensional regularization \cite{T}. Newly, the non-local regularization (NLR) \cite{Kle,Woo} coupled with the field-antifield formalism \cite{Paris,PT,eu} 
has been developed. The success of the NLR is based on its power to compute the anomaly on higher-loops. Three of us, recently, have calculated the one-loop anomaly of the $SU(2)$ Skyrme model, using the NRL formalism \cite{aao}.  

In this work we analyze the symmetries disclosed in the conversion method that are destroyed at the quantum level.  Inside the field-antifield point of view, 
this so-called anomaly (as we said above) is also important because it brings an impossibility to solve the quantum master equation. In the computation of the one-loop anomaly of the $O(N)$ nonlinear sigma model, we use the BV quantization coupled to NLR.

The paper is organized as follows: in section 2
we use the BFFT procedure to describe the $O(N)$
nonlinear sigma model as a gauge theory; in section
3 the quantization of the model following the
Schr\"{o}dinger functional method is accomplished for $N=2$.
With the gauge theory we compute the BRST
transformations and calculate the one loop anomaly
using the above mentioned non-local BV formalism.
This is done in section 4. The final considerations are in section 5.

\section{O(N) nonlinear sigma model: non-Abelian BFFT embedding}
\renewcommand{\theequation}{2.\arabic{equation}}
\setcounter{equation}{0}
The $O(N)$ nonlinear sigma model is described by the Lagrangian density

\begin{equation}
{\cal L}=\frac{1}{2}\,\partial_\mu\phi^a\partial^\mu\phi^a
+\frac{1}{2}\,\lambda\,\bigl(\phi^a\phi^a-1\bigr),
\label{3.1}
\end{equation}
where the $\mu=0,1$ and $a$ is an index related to the $O(N)$
symmetry group. The second class constraints of the theory are 

\begin{eqnarray}
T_1&=&\phi^a\phi^a-1,
\nonumber\\
T_2&=&\phi^a\pi_a.
\label{3.8}
\end{eqnarray}
Following the prescription of the BFFT method with a non-Abelian algebra,
the new first class constraints are given by 

\begin{eqnarray}
&&\tilde T_1=\phi^a\phi^a-1 + \eta^1,
\label{3.22}\\
&&\tilde T_2=\phi^a\pi_a - \eta^2 + \eta^1 \eta^2,
\label{3.23}
\end{eqnarray}
where $\eta^1$ and $\eta^2$ are auxiliary fields that
satisfy the following algebra

\begin{eqnarray}
\label{algebra}
\{ \eta^a, \eta^b \} = 2\delta(x-y).
\end{eqnarray}
The first class constraint algebra is

\begin{eqnarray}
&&\bigl\{\tilde T_1(x),\,\tilde T_1(y)\bigr\}=0,
\nonumber\\
&&\bigl\{\tilde T_1(x),\,\tilde T_2(y)\bigr\}
=2\,\tilde T_1(x)\,\delta(x-y),
\nonumber\\
&&\bigl\{\tilde T_2(x),\,\tilde T_2(y)\bigr\}=0.
\label{3.24}
\end{eqnarray}
Our next step is the calculation of the extended canonical Hamiltonian density. 
The canonical Hamiltonian density is  

\begin{equation}
{\cal H}_c=\frac{1}{2}\,\pi_a\pi_a
-\frac{1}{2}\,\partial_i\phi^a\partial^i\phi^a
-\frac{1}{2}\,\lambda\,\Bigl(\phi^a\phi^a-1\Bigr).
\label{3.25}
\end{equation}
In order to derive the corresponding Hamiltonian in the extended phase space,
we consider\cite{ON} 

\begin{eqnarray}
&&\tilde H_c=\int dx\,\Bigl[\frac{1}{2}\,\pi_a\pi_a(1-\eta^1)
-\phi^a\pi_a\eta^2(1-\eta^1)\nonumber\\
&&+\frac{1}{2}\phi^a\phi^a\eta^2\eta^2(1-\eta^1)
-\frac{1}{2}\,\phi^a\partial_i\partial_i\phi^af(\eta^1)\Bigl],
\label{3.26}
\end{eqnarray}
where $f(\eta^1)$ is an unknown function of the auxiliary field 
$\eta^1$. In order to obtain $f(\eta^1)$, let us demand that

\begin{equation}
\label{3.27}
\{ \tilde{T_\alpha},\tilde{H_c} \} = 0, \,\,\,\, \alpha=1,2.
\end{equation}
We note that this expression is evident for $\alpha=1$. From the equation for $\alpha=2$, we get

\begin{equation}
\label{3.28}
\frac{{f\prime}(\eta^1)}{f(\eta^1)}=\frac{1}{1-\eta^1},
\end{equation}
where the prime means derivative with respect to the auxiliary field $\eta^1$. From Eq.(\ref{3.28}) we have

\begin{equation}
\label{3.29}
f(\eta^1)=\frac{1}{1-\eta^1}.
\end{equation}
Substituting expression (\ref{3.29}) into (\ref{3.26}), we obtain

\begin{eqnarray}
&&\tilde H_c=\int dx\,\Bigl[\frac{1}{2}\,\pi_a\pi_a(1-\eta^1)
-\phi^a\pi_a\eta^2(1-\eta^1)\nonumber\\
&&+\frac{1}{2}\phi^a\phi^a\eta^2\eta^2(1-\eta^1)
-\frac{1}{2}\,\phi^a\partial_i\partial_i\phi^a\frac{1}{1-\eta^1}\Bigl].
\label{3.30}
\end{eqnarray}
In order to embed the $O(N)$ nonlinear sigma model in a non-Abelian gauge theory
we use the equivalent first class Hamiltonian, which differs from the involutive Hamiltonian
(\ref{3.30}) by the addition of a term proportional to the first class constraint $\tilde T_2$, as follows, 

\begin{eqnarray}
&&\tilde {H}_{c_1}=\int dx\,\Bigl[\frac{1}{2}\,\pi_a\pi_a(1-\eta^1)
-\phi^a\pi_a\eta^2(1-\eta^1)\nonumber\\
&&+\frac{1}{2}\phi^a\phi^a\eta^2\eta^2(1-\eta^1)
-\frac{1}{2}\,\phi^a\partial_i\partial_i\phi^a\frac{1}{1-\eta^1}
+\eta^2(\phi^a\pi_a - \eta^2 + \eta^1 \eta^2)
\Bigr].
\label{3.31}
\end{eqnarray}
We note that this Hamiltonian satisfies the first class Poisson algebra

\begin{eqnarray}
&&\bigl\{\tilde T_1,\,\tilde {H}_{c_1}\bigr\}=2\tilde T_2+2\eta^2\tilde T_1,
\nonumber\\
&&\bigl\{\tilde T_2,\,\tilde {H}_{c_1}\bigr\}=2\eta^2\tilde T_2.
\label{3.33}
\end{eqnarray}

Finally, we look for the Lagrangian that leads to this new theory.
A consistent way of doing this is by means of the constrained path integral 
formalism, where the Faddeev procedure \cite{Faddeev} has to be used.

In the Hamiltonian formalism, let us identify the new variables $\eta^a$ as a canonically conjugate pair 
$(\varphi, \pi_\varphi)$,

\begin{equation}
\label{cpair1}
\eta^1 \rightarrow 2 \varphi \,,
\end{equation}
and
\begin{equation}
\label{cpair2}
\eta^2 \rightarrow \pi_\varphi. 
\end{equation}

\noindent They satisfy the relation (\ref{algebra}).
Then, the general expression for the vacuum functional reads

\begin{equation}
\label{vfg}
Z = N \int [d\mu] \exp \{ i \int dx dt [ \dot{\phi}^a\pi_a
+ \dot{\varphi}\pi_\varphi - \tilde{H_{c_1}} ] \},
\end{equation}

\noindent with the measure  $[d\mu]$ given by

\begin{eqnarray}
\label{mesure}
[d\mu] &=& [d\phi^a] [d\pi_a] [d\varphi] [d\pi_\varphi]
| det\{,\} |
\delta(\phi^a \phi^a-1+2\varphi)
\delta(\phi^a\pi_a - \pi_\varphi + 2 \varphi \pi_\varphi) \prod_\alpha
\delta(\tilde{\Lambda}_\alpha)\;\;, \nonumber \\
& &
\end{eqnarray}

\noindent where $\tilde{\Lambda}_\alpha$ are the gauge fixing conditions corresponding to the first class constraints $\tilde{T}_\alpha$ and the term $| det\{,\} |$ represents the determinant of all constraints of the theory, including the gauge-fixing ones.  The quantity $N$ that appears in 
(\ref{vfg}) is the usual normalization factor.  Using the delta functions $\delta(\phi^a\phi^a - 1 
 + 2 \varphi)$, $\delta(\phi^a\pi_a-\pi_\varphi(1-2\varphi))$ and exponentiating the last one with the Fourier variable $\xi$, we obtain

\begin{eqnarray}
\label{ele}
Z &=& N \int [d\phi^a] [d\pi_a] [d\varphi] [d\pi_\varphi] [d\xi]
| det\{,\} | \, \delta( \phi^i\phi^i-1+2\varphi ) \nonumber \\
& &\prod_\alpha
\delta(\tilde{\Lambda}_\alpha)  \exp \{ i \int dx dt 
[  \dot{\phi}^a \pi_a + \dot{\varphi} \pi_\varphi  
-{1 \over 2} \pi_a \pi_a \phi^b \phi^b +
 \pi_a \pi_a \phi^b \phi^b \pi_\varphi \nonumber \\
&-&
{1 \over 2} (\phi^a \phi^a)^2 \pi_\varphi \pi_\varphi +
\xi \phi^a \pi_a - \xi \phi^a \phi^a \pi_\varphi
+\frac{1}{2}\phi^a\partial_i\partial_i\phi^a\frac{1}{1-2\varphi}] \}\;\;.
\end{eqnarray}

\noindent Performing the integration over the momenta and the variable $\xi$, we obtain
\medskip

\begin{eqnarray}
\label{elei3}
Z &=& N \int [d\phi^i] [d\varphi] {1\over (\phi^a\phi^a)^{1/2}} \, 
\delta(\phi^a\phi^a -1+2\varphi) \,\,
\delta(2\phi^a\dot{\phi}^a + 2\dot{\varphi})
\nonumber \\ 
& &\prod_\alpha \delta(\tilde{\Lambda}_\alpha)
 | det\{,\} | 
\exp \{ i \int dx dt [  
{\dot{\phi}^i\dot{\phi}^i\over \phi^i\phi^i}-{1 \over 2}
{\dot{\varphi} \dot{\varphi} \over (\phi^i\phi^i)^2} + \frac{1}{2}\phi^a\partial_i\partial_i\phi^a\frac{1}{1-2\varphi}] \}.
\end{eqnarray}
\medskip

\noindent The new delta function that appears into the expression (\ref{elei3}) was obtained after integration over $\xi$.  We notice that it does not represent any new restriction over the coordinates of the theory and leads to a consistency condition on the constraint 
 $\tilde{T}_1$.  From the vacuum functional (\ref{elei3}), we identify the extended Lagrangian density

\medskip

\begin{eqnarray}
\tilde{\cal L}&=&\frac{1}{2}\frac{\dot\phi^a\dot\phi^a}{1-2\varphi}
+\frac{1}{2}\phi^a\partial_i\partial_i\phi^a\frac{1}{1-2\varphi}
-\frac{1}{2}\frac{\dot\varphi\dot\varphi}{(1-2\varphi)^2}\,\,.
\label{3.32}
\end{eqnarray}

\medskip

In this embedding, with the choice of the non-Abelian algebras (\ref{3.24}) and (\ref{3.33}), we notice that in the expression of the extended Lagrangian (\ref{3.32}) there is not a Liouville term in the auxiliary fields as in the reference \cite{BW}.  The reason for this difference was the choice of another non-Abelian algebra in \cite{BW} .
 
\section{Reduced phase-space quantization in a functional Schr\"{o}dinger representation.}
\renewcommand{\theequation}{3.\arabic{equation}}
\setcounter{equation}{0}
\label{sec:quantization}

In the present section we quantize the $O(N)$ nonlinear sigma model, written as a non-Abelian gauge theory.
We canonically quantize the theory in a functional
Schr\"{o}dinger representation \cite{jac,hatfield,jackiw1}.
Therefore, as the fundamental equations
representing our theory we take the constraints (\ref{3.22})
and (\ref{3.23}) and the Hamiltonian (\ref{3.31}),
all of them written in terms of the conjugated
pair ($\varphi$, $\pi_\varphi$) (\ref{cpair1})
and (\ref{cpair2}).

Due to the presence of the constraints, we have to choose
among the different procedures to canonically quantize a
constrained theory. Here, we use the so-called `reduced
phase-space' quantization \cite{ht}. It means that, we 
have to impose classically the constraints, which reduces
the theory to its physical degrees of freedom. Then, we  
re-write the Hamiltonian ($\tilde{H}_c$) in terms of these physical
degrees of freedom. Finally, we canonically quantize this reduced Hamiltonian ($H_{c}^{r}$) in a functional 
Schr\"{o}dinger representation.

We start imposing, classically, the constraints (\ref{3.22}) and (\ref{3.23}), which gives,

\begin{equation}
\label{m}
\phi^a \phi^a\, -1\, +\, 2\varphi\, =\, 0\, ,
\end{equation}
\begin{equation}
\label{m+1}
\phi^a \pi_a\, -\pi_\varphi\, +\, 2\varphi
\pi_\varphi\, =\, 0\, .
\end{equation}
Now, using the above equations (\ref{m}) and (\ref{m+1}),
we express the field $\varphi$ and its conjugated
momentum $\pi_\varphi$ in terms of the fields $\phi^a$ and their conjugated momenta $\pi_a$. Next, we introduce
the expression relating ($\varphi , \pi_\varphi$) with
($\phi^a , \pi_a$) in the $\tilde{H}_c$ (\ref{3.31}), 
obtaining in this way the following reduced Hamiltonian,

\begin{equation}
\label{m+2}
H_{c}^{r}\, =\, \int \left[ (\pi_a \pi_a )
(\phi^b \phi^b ) \, -\,
(\phi^a \pi_a )
(\phi^b \pi_b ) \, -\,
\phi^a \partial_i \partial_i \phi^a {1\over
\phi^b \phi^b} \right] dx\, .
\end{equation}
It is important to notice that  in the derivation of the
above expression of $H_{c}^{r}$ (\ref{m+2}), from 
$\tilde{H}_c$
(\ref{3.31}), we have also set to zero the last term of
$H_c$. As one can see this term is proportional to the
constraint (\ref{m+1}).

In order to simplify our treatment, we restrict
our attention to the case of two fields ($N=2$).
Therefore, we re-write
$H_{c}^{r}$ (\ref{m+2}) in terms of the fields $\phi_1$ 
and $\phi_2$ and their respective momenta $\pi_1$ and
$\pi_2$, as,

\begin{equation}
\label{m+3}
H_{c}^{r}\, =\, {1\over 2} \int \left[ \left( \pi_1 
\phi_2 - \pi_2 \phi_1 \right)^2\, -\, {1\over 
\phi_{1}^{2} + \phi_{2}^{2}}\left(\phi_1 \partial_i
\partial_i \phi_1 + \phi_2 \partial_i \partial_i \phi_2
\right)\right] dx\, .
\end{equation}

Our next step is the quantization, in the functional
Schr\"{o}dinger representation, of the reduced theory
described by $H_{c}^{r}$ (\ref{m+3}).

We start considering $\phi_1$ , $\phi_2$ , $\pi_1$ and
$\pi_2$ as quantum operators, it means that in the fields basis the momenta is replaced by the following functional derivatives,

\begin{equation}
\label{m+4}
\pi_1 (x)\, \to\, - i { \delta \over \delta \phi_1 (x)}
\quad , \quad \pi_2 (x)\, \to\, - i {\delta \over \delta
\phi_2}\, ,
\end{equation}
where we have set $\hbar$ equal to one.

In general, the states of the theory are given by
time-dependent functionals of the fields, namely,

\begin{equation}
\label{m+5}
\Psi\, =\, \Psi [\phi_1 , \phi_2 , t ]\, .
\end{equation}

This wave-functional $\Psi$ (\ref{m+5}) satisfies the
Schr\"{o}dinger equation,

\begin{equation}
\label{m+6}
i {\partial \over \partial t} \Psi [\phi_1 , \phi_2 , t]\,
=\, \hat{H}_{c}^{r} [\phi_1 , \phi_2 , t]\, ,
\end{equation}
which is a functional differential equation because 
$\hat{H}_{c}^{r}$, with the aid of (\ref{m+3}) and 
(\ref{m+4}), is given by,

\begin{equation}
\label{m+7}
\hat{H}_{c}^{r}\, =\, {1\over 2} \int \left[ \left( -i {
\delta \over \delta \phi_1} \phi_2 + i {\delta
\over \delta \phi_2} \phi_1 \right)^2\, -\, {1\over
\phi_{1}^{2} + \phi_{2}^{2}}\left( \phi_1 \partial_i
\partial_i \phi_1 + \phi_2 \partial_i \partial_i \phi_2
\right) \right] dx\, .
\end{equation}

Observing the $O(2)$ symmetry of our sigma nonlinear 
model, it seems interesting, to re-write $\hat{H}_{c}^{r}$
in terms of a new pair of fields $(R , \Theta )$, related to the old ones $( \phi_1 , \phi_2 )$ , by,

\begin{equation}
\label{m+8}
\phi_1\, =\, R \sin \Theta \quad \mbox{and} \quad \phi_2\,
=\, R \cos \Theta \, .
\end{equation}

In terms of the new fields $(R , \Theta)$ and their 
respective functional derivatives, which may be derived 
from (\ref{m+8}), $\hat{H}_{c}^{r}$ is written as,

\begin{equation}
\label{m+9}
\hat{H}_{c}^{r}\, =\, {1\over 2} \int \left[ -\, {\delta^2
\over \delta \Theta^2}\, -\, \Theta \partial_i \partial_i
\Theta \, -\, {1\over R} \partial_i \partial_i R \right] dx
\, .
\end{equation}

It is important to mention that we have solved the 
factor-ordering ambiguities in $\hat{H}_{c}^{r}$ (\ref{m+9}),
by using the so-called `symmetric factor-ordering' 
\cite{lee}.

Since $\hat{H}_{c}^{r}$ does not explicitly depend on time,
we may separate out the time dependence of the 
wave-functional, now given in terms of $R$ and $\Theta$
($\Psi [R, \Theta ,t]$), and write,

\begin{equation}
\label{m+10}
\Psi [R, \Theta ,t]\, =\, e^{-iEt} \Psi [R, \Theta ]\, .
\end{equation}

$\Psi [R, \Theta ]$ satisfies the time-independent
Schr\"{o}dinger functional equation,

\begin{equation}
\label{m+11}
\int \left[ -\, {\delta^2 \Psi \over \delta \Theta^2}\,
-\, \Theta \partial_i \partial_i \Theta \Psi\, -\,
{1\over R} \partial_i \partial_i R \Psi \right] dx\, =\,
2 E \Psi\, .
\end{equation}

Following \cite{hatfield}, in order to find the ground 
state or vacuum wave-functional, $\Psi_0 [R, \Theta ]$, 
we write the following ansatz for $\Psi_0 [R,
\Theta ]$,

\begin{equation}
\label{m+12}
\Psi_0 [R, \Theta ]\, =\, \eta \exp\{-G[R, \Theta ]\}\, .
\end{equation}
Introducing the ansatz (\ref{m+12}) in equation 
(\ref{m+11}), we obtain the below equation for 
$G[R, \Theta ]$,

\begin{equation}
\label{m+13}
\int \left[ {\delta^2 G\over \delta \Theta^2}\, -\,
\left( {\delta G\over \delta \Theta}\right)^2\, -\,
\Theta \partial_i \partial_i \Theta\, -\, {1\over R}
\partial_i \partial_i R \right] dx\, =\, 2 E\, .
\end{equation}

This equation naturally separates in two parts, one that
depends solely on $R$ and another that depends on $R$
and $\Theta$, through $G[R, \Theta ]$. Such that we may
re-write (\ref{m+13}) as,

\begin{equation}
\label{m+14}
\int \alpha (x) dx\, +\, \int \beta (x) dx\, =\, 2 E\, ,
\end{equation}
where,
\begin{equation}
\label{m+15}
-\, {1\over R} \partial_i \partial_i R\, =\, \beta (x)\, ,
\end{equation}
and
\begin{equation}
\label{m+16}
\int \left[ {\delta^2 G\over \delta \Theta^2}\, -\, \left(
{\delta G\over \delta \Theta}\right)^2\, -\, \Theta 
\partial_i \partial_i \Theta \right] dx\, =\, \int 
\alpha (x) dx\, .
\end{equation}

Equation (\ref{m+14}) may be interpreted as saying that the
energy of the system is divided in two parts. The first 
part $\int \beta (x) dx$ is entirely determined by $R$ from
(\ref{m+15}), we call it $E_R$. The second one $\int
\alpha (x) dx$ is determined by a functional that might
depend on $R$ and $\Theta$, we call it $E_\Theta$.

For a given function $\beta (x)$ one finds, with the aid of
(\ref{m+15}) and appropriated boundary conditions upon $R(x)$,
one function $R(x)$. Therefore, we may see that $R(x)$ will
not be allowed to be a generic function in the function space. 

It is important to notice that $\beta(x)$ has to 
satisfy certain conditions such as finiteness and 
positiveness of $\int \beta (x) dx$. For a positive $\int
\beta (x) dx$, we may define the ground sate of $E_R$ to be
the one where $\beta (x) = 0$. States with $\beta (x) \neq 0$
would represent the excited states.

Equation (\ref{m+16}) is well-known in the literature of
quantization in a functional Schr\"{o}dinger representation.
It is the equation for a massless scalar field \cite{hatfield}.
The functional $G[R, \Theta ]$ which satisfies (\ref{m+15}),
for the present situation, has the following expression,

\begin{equation}
\label{m+17}
G[R, \Theta ]\, =\, \int dy dx\, \Theta(y) g(y,z) \Theta(z)\,
+\, \int dz \left( - {1\over R(z)} \partial_i \partial_i R(z)
\right)\, ,
\end{equation}
where the last term on the right hand side is simply $\int 
\beta (z) dz$.

Introducing (\ref{m+17}) in (\ref{m+16}), we may obtain the explicit expression for $g(y,z)$,

\begin{equation}
\label{m+18}
g(y,z)\, =\, {1\over 2} \int {dk\over 2\pi} k e^{ik(y-z)}\, ,
\end{equation}
and the ground state energy of $E_\Theta$. This energy is derived by computing $\int \alpha (x) dx$, which gives,

\begin{equation}
\label{m+19}
\int \alpha (x) dx\, =\, \int g(x,x) dx\, =\, {1\over 2}
\int {dk\over 2\pi} k \int dx\, =\, {1\over 2} \int dk k
\delta (0)\, .
\end{equation}
It agrees with the result obtained in the operator 
representation \cite{hatfield}.

An important result comes from (\ref{m+19}). $G[R, \Theta ]$(\ref{m+17}), has two components: one that depends on $\Theta$ ($G_\Theta$) and another that depends on $R$ ($G_R$). From (\ref{m+19}), it is clear that $E_\Theta$, $\int \alpha
(x) dx$, is entirely determined by $G_\Theta$. Therefore it does not depend upon $R$.

Now, we set $\beta (x) = 0$ in (\ref{m+17}), accordingly to our suggestion to the ground state energy of $E_R$, and combine the resulting expression with (\ref{m+12}). Then, we may obtain the normalized ground state wave-functional of
the Fourier transform of $\Theta (x)$ ($\Psi_0 [\tilde{\Theta} 
(k)]$) as \cite{hatfield},

\begin{equation}
\label{m+20}
\Psi_0 [\tilde{\Theta} (k)]\, =\, \prod_k \left({k\over \pi}
\right)^{1/4} \exp\left({-1\over 4\pi} k \tilde{\Theta}^2 (k)
\right)\, .
\end{equation}
This is just the infinite product of harmonic oscillators
ground state wave-functions, one wave-function for each
$k$.

For the excited states we have $\beta (x) \neq 0$
which, from (\ref{m+17}), would introduce a $R$ dependence
in the wave-functionals. As a typical example we may write,

\begin{equation}
\label{m+21}
\Psi_1 [R, \Theta ]\, =\, {k_1\over \pi}^{1/2} \int dy 
e^{-ik_1 y} \Theta (y) \Psi_0 [\Theta ] \exp \left[
\int dz {1\over R(z)} \partial_i \partial_i R(z) \right]
\, .
\end{equation}
It represents the wave-functional associated with the first
excited state of $E_\Theta$ \cite{hatfield}, and a generic excited state of $E_R$.

As a matter of completeness we would like to mention the reference \cite{kim}, where the functional Schr\"odinger representation was first applied to the study of $O(N)$ nonlinear sigma models.  We may identify two main differences between \cite{kim} and the present work.  Firstly, we have re-written the theory as a non-Abelian gauge theory, and in \cite{kim} it was treated as a second-class system.  Secondly, we have explicitly solved the functional Schr\"odinger equation and found the ground state as well as the excited wave-functionals for $N=2$.  In \cite{kim} it was computed the expected value of the Hamiltonian using a trial wave-functional, in the large $N$ limit.

\section{The one-loop anomaly}
\renewcommand{\theequation}{4.\arabic{equation}}
\setcounter{equation}{0}

In this section we follow the standard references about the BV formalism \cite{BV,GPS,Jon,Hen} and the NLR \cite{Kle,Woo} coupled to the field-antifield procedure \cite{ABD,Paris,PT,eu}.  All the details of the theory involved in the following calculation of the $O(N)$ nonlinear sigma model anomaly can be found in those papers.

The first class constraint (\ref{3.22}), written in terms of $\varphi$ of equation (\ref{cpair1}), tell us that 

\begin{equation}
\varphi=1-\phi^a \phi^a\;\;,
\end{equation}
so that, 

\begin{equation}
\dot{\varphi}\,=\,-\phi^a \dot{\phi}^a\;\;.
\end{equation}

Substituting this in (\ref{3.32}) we have now that

\begin{equation}
\label{action2}
{\cal S}\,=\,{1 \over 2}\int\,dx\,
dt\,\left[\frac{\dot\phi^a\dot\phi^a\,+\,\phi^a\partial_i\partial_i\phi^a}{\phi^a\,\phi^a}
\,-\,\frac{(\,\dot\phi^a\dot\phi^a\,)^2} {(\,\phi^a\,\phi^a\,)^2}\right]\;\;,
\end{equation}

\noindent This action, as we can easily see, has a problem of non-locality, which can be solved 
expanding the terms,

\begin{eqnarray}
\label{action3}
{\cal S}\,&=&\,{1 \over 2}\,\int\,
dt\,\left\{\frac{\dot\phi^a\dot\phi^a\,+\,\phi^a\partial_i\partial_i\phi^a}
{[1\,-\,(\,1\,-\,\phi^a\,\phi^a\,)]}
\,-\,\frac{(\,\dot\phi^a\dot\phi^a\,)^2} {[1\,-\,(\,1\,-\,\phi^a\,\phi^a\,)]^2}\right\} 
\nonumber \\
&=& \,{1 \over 2}\,(\,\dot\phi^a\dot\phi^a\,+\,\phi^a\partial_i\partial_i\phi^a\,) \sum_{n=0}^\infty (1-\phi^a\,\phi^a)^n\, \nonumber \\
&-& \, {1 \over 2}\,(\,\dot\phi^a\dot\phi^a\,)^2 \sum_{n=0}^\infty (n+1)(1-\phi^a\,\phi^a)^n\;\;.
\end{eqnarray}

\noindent After a simple calculation, we can say that 
this action is invariant under the BRST transformations given by 

\begin{equation}
\delta
\phi^a \, = \,c\,\phi^a\;,
\end{equation}
and
\begin{equation}
\delta c\,= \, 0\;\;,
\end{equation}
where $c=c^a\,T^a$ and $tr\,(T^a\,T^a)={1 \over 2}$\,.

Now we can construct the BV action,

\begin{eqnarray}
\label{actionBV}
{\cal S}_{BV}\,&=& \,{1 \over 2}\,  \int dt\,dx\, \left\{\,(\,\dot\phi^a\dot\phi^a\,+\,
\phi^a\partial_i\partial_i\phi^a\,)  \sum_{n=0}^\infty 
(1-\phi^a\,\phi^a)^n\, \right.\nonumber \\
&-& \left.\,(\,\dot\phi^a\dot\phi^a\,)^2 \sum_{n=0}^\infty (n+1)(1-\phi^a\,\phi^a)^n 
\,+\,{\phi^a}^* \,c\, \phi^a\, \right\}.
\end{eqnarray}

\noindent The kinetic part of the action (\ref{action2}) (i.e., with $n=0$) after an integration by parts is,

\begin{eqnarray}
F  &=& \,{1 \over 2}\,\int d\,t\,[\,\dot{\phi}^a\,\dot{\phi}^a\,] \nonumber \\
&=& {1 \over 2}\,\int dt [\,-\,\phi^a\,\p^2_0\, \phi^a\,]\;\;.
\end{eqnarray}

\noindent Hence, the kinetic term has the form

\begin{eqnarray}
F &=& \frac{1}{2} \phi^a (-\, \p^2_0 )\, \phi^a\;\; \nonumber \\
&\Longrightarrow& \;\; {\cal F}_{AB}\,=\,
- \,\p^2_0 \,\delta_{AB}.
\end{eqnarray}

The regulator, a second order differential operator, can be chosen as

\begin{eqnarray}
{\cal R}^A_B &=& \p^2_0\,\delta^A_{\;\;\;B} \nonumber \\
&\Longrightarrow&\;\;T\,=\,-\,1\;\;.
\end{eqnarray}

\noindent where $T$, as required, is clearly an arbitrary non-singular matrix.

The smearing operator has the form,

$$\epsilon^{A}_{\;\;\;B} = exp \left( \frac{\partial^{2}_0}{
2\,\Lambda^{2}} \right)
\,\delta^A_{\;\;\;B}\;\;.$$

In the NLR scheme the shadow kinetic operator is

\begin{equation}
{\cal O}_{AB}^{-1} \,=\,\left( \frac{{\cal F}}{\epsilon^{2} - 1}
\right)_{AB}
= \left( \frac{-\,\p^2_0}{\epsilon^{2} - 1} \right)_{AB},
\end{equation}

\noindent where

\begin{equation}
{\cal O}^{AB} = -\,\frac{\epsilon^{2} - 1}{\p^2_0}
\;=\;-\,\int^1_0 \frac{d \tau}{\Lambda^2}\, exp \left( \tau \frac{\p^2_0}{\Lambda^2}\right).
\end{equation}

\noindent Using the definitions of $S^A_{\;\;B}$ and ${\cal I}_{AB}$, we can show that
\begin{eqnarray}
\label{fla}
S^\phi_{\;\;\phi}\,&=&\,c,\\
{\cal I}_{\phi \phi} &=& -\, \p^2_0\;+\;\frac{\p^2_0\,+\,\p^2_i}{\phi^a\,\phi^a}\;-\;
\frac{4 \phi^a \dot{\phi}^a \p_0 \;+\;3\,\phi^a\,\p_i^2\,\phi^a\,+\,2\,\phi^a\phi^a\,\p_i^2
\,+\,\dot{\phi}^a \dot{\phi}^a\,+\,\p_0}{(\phi^a\,\phi^a)^2} \nonumber \\
&+&\frac{2\,\phi^a\,\phi^a\,(\,\dot\phi^a\dot\phi^a\,+\,\phi^a\partial_i\partial_i\phi^a\,)
\,+\,3\,\phi^a\,\dot{\phi}^a  
\,+\,2\,\phi^a\,\phi^a\,\p_0}{(\phi^a\,\phi^a)^3}  \nonumber \\
&-& \, 2 \,\frac{\phi^a\,\phi^a\,(\phi^a\, \dot{\phi}^a)}{(\phi^a\,\phi^a)^4}.
\end{eqnarray}

\noindent Finally, the anomaly can be computed as we know

\begin{eqnarray}
{\cal A}\,&=&\,(\Delta S)_R\, \nonumber \\
&=&\,\lim_{\Lambda^2 \rightarrow \infty} \{\,Tr[\,\epsilon^2 \,S^A_{\;\;B}\,]
\,+\,Tr[\,\epsilon^2\, S^A_{\;\;D}\, {\cal O}^{DC}\,{\cal I}_{CB}\,]\, \}.
\end{eqnarray}

\noindent Computing each term, we have that the only non-zero integral is the one coming from the second term 
in ${\cal I}_{\phi\phi}$ (\ref{fla}).  Note that, differently from \cite{aao}, now we are in a two dimensional space.  Let us write  only the main steps of this calculation. 
So,

\begin{eqnarray}
& &
\!\!\!\!\!\!\!\!\!\!\!\!\!\!\!\!\!\!\!\!\lim_{\Lambda^2 \rightarrow \infty} 
\left[ \,\epsilon^2\,c\,\int\,dt\,dx \,{\cal O}
\, \frac{(\p^2_0\,+\,\p^2_i)}{\phi^a \phi^a}\right] \nonumber\\
&=& 
\lim_{\Lambda^2 \rightarrow \infty} \[ \,\epsilon^2\,c\,\int\,dt\,dx \int 
\frac{d^2 k}{(2 \pi)^2} e^{-ikx}\,{\cal O} \frac{(\p^2_0\,+\,\p^2_i)}{\phi^a \phi^a}
exp \left( \frac{\p^2}{\Lambda^2} \right) 
e^{ikx} \] \nonumber \\
&=& 
\lim_{\Lambda^2 \rightarrow \infty} \[ \,\epsilon^2\,c\,\int\,dt\,dx\,
\int \,\frac{d^2 k}{(2 \pi)^2} e^{-ikx}\,  
\,\int^1_0 \(-\, \frac{d \tau}{\Lambda^2} \)\, exp 
\left( \tau \frac{\p^2_0}{\Lambda^2}\right)\, \frac{(\p^2_0\,+\,\p^2_i)}{\phi^a \phi^a}
exp \left( \frac{\p^2}{\Lambda^2} \right) e^{i\,k\,x}\] \nonumber \\
&=& 
\lim_{\Lambda^2 \rightarrow \infty} \[ \,\epsilon^2\,c\,\int\,dt\,dx\, 
\,\int^1_0 \(- \frac{d \tau}{\Lambda^2} \)\, exp 
\left( \tau \frac{\p^2_0}{\Lambda^2}\right)\, \frac{1}{\phi^a \phi^a} \right. \nonumber \\
&\qquad\,\times& \left.
\int \,\frac{d\,k_0\,d\,k_1}{(2 \pi)^2} \, (-\,k_0^2\,-\,k_1^2)\,
exp \left( \frac{\,-\,k_0^2\,+\,k_1^2\,}{\Lambda^2} \right) \] \nonumber \\
&=& 
\lim_{\Lambda^2 \rightarrow \infty} \[ \,\frac{\pi\,c}{4}\,\epsilon^2\,\int\,dt\,dx\,
\int^1_0 d \tau \left( 1 + \tau\,\frac{\p^2_0}{\Lambda^4} \right)
\frac{1}{\phi^a \phi^a}\] \nonumber \\
&=& 
\frac{\pi\,}{4} \, \int\,dt\,dx \,\frac{c}{\phi^a \phi^a}, 
\end{eqnarray}

\noindent where we make two convenient reparametrizations, 

\begin{equation}
(\tau,k) \rightarrow (\frac{\tau}{\lambda^2},\lambda k),
\end{equation}

\noindent to solve the integrals \cite{grad}.

Repeating the same procedure (integration) for all the other terms of the $\cal{I_{\phi \phi}}$ (\ref{fla}) one can conclude that they are identically zero, as we have said above.

As we know, terms that depend only on ghosts do not have any physical meaning in the final 
result of the anomaly.  Computing only the physical terms, the one-loop anomaly for the $O(N)$ nonlinear sigma model is the Wess-Zumino consistent expression \cite{WZ},

\begin{equation}
{\cal A}\, = \,\frac{\pi}{4}
\int\,dt \, {c \over \phi^a\,\phi^a}\;\;.
\end{equation}

\noindent It is a new result, showing that, at the level of the BV formalism,  the QME has no solution.

\section{Conclusions}

In this work we studied the $O(N)$ nonlinear sigma model embedding this system in a non-Abelian gauge theory.  It was accomplished through an extension of the BFFT conversion method. 

Following the prescription of this method, we obtained the new first-class constraints, the extended Hamiltonian and the effective Lagrangian that leads to the new theory. 

Then, we quantized the first class O(2) nonlinear sigma model using the functional Schr\"{o}dinger representation. Since this theory is constrained, we applied the so-called ``reduced phase-space" quantization procedure. The crucial step, in this program, was the polar transformation
from the original fields $(\phi_1,\phi_2)$ to new
fields $(R,\Theta)$. This transformation is naturally
suggested by the O(2) symmetry of the theory. In terms
of $(R,\Theta)$, the functional Schr\"{o}dinger equation was greatly simplified. From this equation it was clear that the energy of the theory is divided in two parts: a radial one (depends only on R) and an angular one (depends only on $\Theta$). With an appropriated suggestion for the ground state energy, we explicitly computed the ground state wave-functional
and indicated how to calculate the excited
states wave-functionals.

Finally, we have computed the anomaly at one-loop order of the $O(N)$ nonlinear sigma model through the introduction of its BRST transformations and consequently of all the ingredients of the field-antifield procedure. 

The importance of a conversion constraint method is fundamentally to have a gauge theory. These gauge symmetries, at the classical level, give rise to conserved Noether currents. In many cases, including
the one studied here, the expected value of the Noether currents are not conserved. 

Based on the results obtained in \cite{aao} and in this
work, we believe that the BFFT procedure of extension of the phase space affects the Wess-Zumino sector of the O(N) non-linear sigma model. At the quantum level, it causes the non conservation of the Noether currents
generated by the classical gauge symmetries mentioned above.

\section{Acknowledgments}

The authors would like to thank {\bf FAPEMIG} and {\bf FAPESP}, Brazilian agencies, for the partial financial support of this work.

\end{document}